\documentstyle[preprint,aps]{revtex}
\tightenlines

\begin{document}
\newcommand{\beq}{\begin{equation}}
\newcommand{\eeq}{\end{equation}}
\newcommand{\beqa}{\begin{eqnarray}}
\newcommand{\eeqa}{\end{eqnarray}}
\newcommand{\sr}{\sqrt}
\newcommand{\fr}{\frac}
\newcommand{\mn}{\mu \nu}
\newcommand{\G}{\Gamma}

\draft
\preprint{ INJE-TP-00-06, hep-th/0009117}
\title{Localization of graviphoton and graviscalar   \\
       on the brane}
\author{  Y.S. Myung\footnote{E-mail address:
ysmyung@physics.inje.ac.kr} }
\address{
Department of Physics, Graduate School, Inje University,
Kimhae 621-749, Korea}
\maketitle
\begin{abstract}

The question of whether  the Kaluza-Klein (KK) graviphoton $h_{5\mu}$ and
graviscalar $h_{55}$ are localized or not on the brane is one of
the important issues. In this letter, we address this problem in five dimensions.
Here we consider  the massless (zero-mode) propagations without requiring
the Z$_2$-symmetry on $h_{5\mu}$.
We obtain  the graviton $h_{\mu\nu}$, graviphoton,  and graviscalar exchange
amplitudes on shell.
We find that the graviscalar has a tachyonic mass. It turns out
that $h_{5\mu}$ admits the localized zero-modes on the brane while $h_{55}$  does not
have a localized zero-mode. This is contrasted to the fact that
the bulk spin-0 field has a localized zero-mode on the brane but the bulk
spin-1 field does not have a localized solution in five dimensions.

\end{abstract}
\bigskip

\newpage

In the conventional  Kaluza-Klein (KK) approach of the five dimensional
(5D) gravity,
the spacetime manifold is factorized as $M_4 \otimes
S^1$. Here $M_4$ is the Minkowski spacetime and $S^1$ is the circle.
The spectrum of the
5D pure gravity is split into   four dimensional (4D) massless
fields such as graviton,  graviphoton,  graviscalar, and an
infinite tower of massive spin-two fields~\cite{CZ,SSRSS}.
In particular, the U(1) gauge symmetry of the graviphoton  in
the 4D effective action is originated from the translational isometries
in the extra dimension.

On the other hand, there have been much interest in the phenomenon of
localization of gravity proposed by Randall and Sundrum
(RS)~\cite{RS2,RS1}. There have been developed
a large number of brane world models afterwards~\cite{CEHS,OMs,GG}.
RS assumed
a single positive tension 3-brane and a negative bulk cosmological
constant in the 5D spacetime~\cite{RS1}. They have obtained a 4D localized gravity
by fine-tuning the tension of the brane to the cosmological constant.
The introduction of branes usually gives rise to the warping of the
extra dimensions, resulting in  the non-factorizable spacetime.
Apparently, the presence of the brane breaks the translational
isometries in the extra dimensions.
Hence we worry about obtaining the graviphoton in the RS approach.

It is not easy to derive the zero-mode propagations
in the non-factorizable compactification. In the conventional KK
approach, one can  find the zero mode $f(x)$ only by requiring
$\partial_5f=0$ in the equation of motion.
 This is possible  because it is in the factorizable compactification.
In this case  the 5D Laplacian is split
into the 4D Laplacian $\Box$  and $\partial_5^2$,
 the latter produces a
(mass)$^2$-term~\cite{SSRSS}. Equivalently the 4D effective action for
zero-modes can be obtained after the integration  of
the 5D pure gravity action over $x^5=z \in S^1$. However, in the
non-factorizable compactification, there exist additional terms that are
function of $x^5=z \in R$ in the linearized equation. In general  we cannot
obtain the consistent zero-mode solution   only by requiring
$\partial_5f=0$ in the level of the equation of motion.
The integration of the  5D  action over $z$ is a  good starting point
to obtain the zero-mode solution for the non-factorizable
compactification~\cite{Chiba,BG,Youm,ML}.
Also the study of this issue is very important for the phenomenological
purposes because its zero modes (massless modes) correspond
to the standard model particles localized on the brane.
If the brane world scenario is correct, the various fields we
observe  are the zero-modes of the KK \cite{MK3} and  bulks fields
\cite{Oda2} which are trapped on the brane by the gravitational interaction
\footnote{The field theoretic
mechanism for gauge field localization on a brane was first suggested in\cite{GMS}.
Also the localization of quantum fields on the brane was recently
discussed\cite{MP}.}.
Here we consider the KK fields only.

A simple choice for the KK fields is a part $(h_{5\mu}=h_{55}=0)$ of
 the RS  gauge\footnote{This is composed of the Gaussian-Normal
 gauge : $(h_{5\mu}=h_{55}=0)$ and the 4D tansverse traceless
 gauge : $\partial_\mu h^{\mn}=0, h^\mu_\mu(h)=0$.}
 to obtain the localized 4D gravity on the brane~\cite{RS1}.
Th brane-bending effect appears  under the RS gauge
with the localized matter source~\cite{GT}.
 Here  the bending of the wall  $\hat\xi^5$  acts like
  a new scalar under the RS gauge\footnote{For the other approach, see ref.\cite{AIM}
  .}.
Ivanov and Volovich~\cite{IV} and Myung and Kang~\cite{MK1}
 have  discussed  the propagation of $h_{5\mu}$ with $h_{55}=0$.
 The propagation of all metric
components including  $h_{5\mu}\not=0, h_{55}\not=0$ was investigated in~\cite{MK2}.
It turned out that the massive modes of $h_{5\mu}, h_{55}$  with uniform external
sources cannot  propagate on the branes.
Recently, the case of $h_{55}\not=0, h_{5\mu}=0$ with the localized source
was discussed~\cite{Kak}. However, it was pointed out that at long distance where
we can obtain the 4D gravity, the propagation of $h_{55}$ is not allowed.
 The next question is
whether the massless modes of graviphoton $h_{5\mu}$ and
graviscalar $h_{55}$ propagate or not on the brane.

It is known for
$h_{55}\sim \varphi$ on the  RS brane  that  the consistency with  $h^\rho_\rho=h_{5\mu}=0$
requires $h_{55}=0$ without the external source~\cite{ML}.
Concerning the massless propagation of $h_{5\mu}\sim a_\mu$,
we expect that the breaking of isometries in the extra dimension by the
brane makes the 4D effective action not being invariant under U(1) gauge
transformations manifestly~\cite{RS2}.
Explicitly, it comes from the Z$_2$-symmetry
argument.
This is based on the fact that RS ground state solution is symmetric under $z \to -z$.
If one requires that this symmetry be preserved up to the linearized
level,
 $h_{\mn}(x,z),h_{55}(x,z)$
are even with respect to $z$, but $h_{5\mu}$   is odd
: $h_{5\mu}(x,-z)=-h_{5\mu}(x,z)$.
This implies that $h_{5\mu}(x,0)=0$ on the brane. Thus we do
not expect to have the zero-modes of the gravivector. Here
we do not require such a Z$_2$-symmetry on the linearized calculation.
Then the analysis of the linearized
equation around the RS background reveals  that the graviphoton
possesses the U(1) gauge symmetry~\cite{MK3}.

In this paper, we clarify  whether the  graviphoton $h_{5\mu}$ and
graviscalar $h_{55}$ are localized or not on the brane with the matter sources.
This  is one of
the important issues about the brane world scenario
~\cite{BG,Chiba,GS,DRT,Oda1}. The naive condition that the
zero-mode  is localized on the brane is equivalent to the
normalizability of the ground state wave function on the
brane~\cite{BG}. This requires that after the integration of the 5D action over $z$
it be finite~\cite{BG,Oda2}. However, this is valid for  the bulk
fields. Actually a further work is necessary for a complete study of
the massless propagations including the graviphoton and  gravisacalar on the brane.
As a definite criterion,
we introduce the local sources to calculate the  graviton, graviphoton,
  and graviscalar exchange
amplitudes on shell.

We start from the second  RS model with a single brane at $z=0$~\cite{RS1,MK3}
\beq
I = \int d^4x \int^{\infty}_{-\infty} dz
    \fr{\sr{-\hat{g}}}{16\pi G_5} (\hat{R} -2\Lambda )
    - \int d^4x \sr{-\hat{g}_B} \sigma .
\label{5DI}
\eeq
Here $G_5$ is the 5D Newtonian  constant, $\Lambda$ is the bulk cosmological
constant, $\hat{g}_B$ is the determinant of the induced
metric for  the 3-brane, and $\sigma$ is the tension of the brane.
We assume that the value of $\sigma$ is fine-tunned such that
$\Lambda =-6k^2 (< 0)$ with $k=4\pi G_5 \sigma /3$. Let us introduce
the domain-wall metric,
\beqa
ds^2 &=& \hat{g}_{MN} dx^Mdx^N = H^{-2}(z) g_{MN} dx^Mdx^N
\nonumber \\
&=& H^{-2}(z) \Big[ \gamma_{\mu\nu}dx^{\mu}dx^{\nu}
    +\Phi^2 (dz -\kappa A_{\mu}dx^{\mu})^2 \Big] .
\label{metric}
\eeqa
Here $H= k|z|+1$, $\Phi^2 = g_{55}$, and $-\kappa \Phi^2 A_{\mu}=g_{5\mu}$.
The above corresponds to the standard KK  decomposition
as
\beq
g_{MN} = \left(\begin{array}{cc}
\gamma_{\mu\nu}+\kappa^2 \Phi^2 A_{\mu}A_{\nu} & -\kappa \Phi^2 A_{\mu} \\
-\kappa \Phi^2 A_{\nu} & \Phi^2
\end{array}\right)
\label{KKm}
\eeq
with $A^{\mu} = \gamma^{\mu\nu} A_{\nu}$ and $A\cdot A = A_{\mu}A^{\mu}$.
Here $\kappa$ is introduced for the  small gauge coupling constant.

Under the specific class of coordinate transformations such as
\beq
x^{\mu} \rightarrow \tilde{x}^{\mu} =  \tilde{x}^{\mu} (x), \qquad
z \rightarrow \tilde{z} = z + \xi(x),
\label{CT}
\eeq
we obtain from $\tilde{g}_{MN} = \fr{\partial x^P}{\partial
\tilde{x}^M}\fr{\partial x^Q}{\partial \tilde{x}^N} g_{PQ}$ as
\beq
\tilde{\gamma}_{\mu\nu} = \fr{\partial x^{\alpha}}{\partial
\tilde{x}^{\mu}} \fr{\partial x^{\beta}}{\partial \tilde{x}^{\nu}}
\gamma_{\alpha\beta}, \qquad \tilde{A}_{\mu} = \fr{\partial
x^{\alpha}}{\partial \tilde{x}^{\mu}} A_{\alpha} + \kappa^{-1}
\fr{\partial \xi}{\partial \tilde{x}^{\mu}},
\qquad \tilde{\Phi} (\tilde{x}, \tilde{z}) = \Phi (x,z)
\eeq
We observe  that $\gamma_{\mu\nu}$ transforms like a 4D
metric tensor and $\Phi$ a scalar field under diffeomorphisms in
Eq.~(\ref{CT}). Also we point out that the 5D diffeomorphisms are
split into the 4D diffeomorphisms plus the U(1) gauge transformations for
$A_{\mu}$.

In this work, we are mainly interested in the zero mode effective
action. It is a non-trivial problem to
determine what the zero mode  is if the full spacetime is not
factorizable. In order to obtain the zero modes, we assume that
$\gamma_{\mu\nu}$, $A_{\mu}$, and $\Phi$ are functions of
$x$-coordinates only.  Plugging Eqs.(2) and (3) with
$\gamma_{\mu\nu}(x)$, $A_{\mu}(x)$, and $\Phi(x)$ into Eq.(1) and integrating
it over $z$ leads to~\cite{MK3}

\beqa
I_{zero-m} &=& \fr{1}{16\pi G_4} \int d^4x \sr{-\gamma}\Big[ \Phi
  R(\gamma ) -\fr{\kappa^2}{4} \Phi^{3} F^2  \nonumber  \\
&& \qquad\qquad\qquad\qquad +6k^2 \Big( \Phi^{-1} +\Phi
  -2\sr{|\delta^{\mu}_{\nu} +\kappa^2 \Phi^2 A^{\mu}A_{\nu}|}
  +\kappa^2\Phi A\cdot A \Big) \Big] .
\label{KKI1}
\eeqa

We observe that the zero-mode gravitational degrees of freedom
in the 5D spacetime are split into the 4D graviton
$\gamma_{\mu\nu}$, a graviphoton $A_{\mu}$, and a graviscalar
$\Phi$ as usual. However, the properties of the
vector field and the scalar field are very different from those
in the conventional KK reduction. The first two terms in Eq.~(\ref{KKI1})
 are the same form as in the conventional
KK compactification, and  thus they have the U(1)
gauge symmetry. The difference from the conventional KK approach
is  the last term which is proportional to $k^2$.
 If we start from the KK metric decomposition with
$A_{\mu}=0$ and $\Phi =1$ in Eq.~(\ref{KKm}) as in the RS approach, this ``potential"
term disappears and one obtains the pure 4D gravity  without the  cosmological constant
on the brane.
The zero cosmological constant arises  because of
the fine-tuning between the brane tension $\sigma$ and the 5D bulk
cosmological constant $\Lambda$.

Apparently the non-linear
term ($\sr{|\delta^{\mu}_{\nu} +\kappa^2 \Phi^2
A^{\mu}A_{\nu}|}$) as well as the squared term  $A\cdot A$
imply
not only that the 4D effective action no longer has the U(1) gauge
symmetry, but also that  the graviphoton does not exist.
 This phenomena arises mainly from
the presence of the brane  in the 5D
Anti de Sitter spacetime. However, this observation is not a complete one.
Actually  the propagation of fields should be determined by the
perturbation analysis around the RS solution.
As will see later,  the non-linear term and $A\cdot A$ cannot
generate any mass-like term.

In order to see explicitly how the dynamical aspect of $\Phi$
 comes out, let us conformally transform the metric as
\beq
\gamma_{\mu\nu} \rightarrow \bar{\gamma}_{\mu\nu} =\Phi \gamma_{\mu\nu}.
\eeq
The zero-mode effective action Eq.~(\ref{KKI1}) is then written by
\beqa
I_{zero-m}^E &=& \fr{1}{16\pi G_4} \int d^4x \sr{-\bar{\gamma}}\Big[
  R(\bar{\gamma}) -\fr{\kappa^2}{4} \Phi^{3} F^2
  -\fr{3}{2} \Phi^{-2} \bar{\nabla}^{\mu}\Phi \bar{\nabla}_{\mu}\Phi
   \nonumber  \\
&& \qquad\qquad  +6k^2 \Phi^{-2}\Big( \Phi^{-1} +\Phi
  -2\sr{|\delta^{\mu}_{\nu} +\kappa^2 \Phi^3 A^{\mu}A_{\nu}|}
  +\kappa^2\Phi^2 A\cdot A  \Big) \Big]
\label{KKI2}
\eeqa
in the Einstein frame. Here  all contractions are done using the metric
$\bar{\gamma}^{\mu\nu}$ as $F^2= \bar{\gamma}^{\mu\nu}
\bar{\gamma}^{\alpha\beta} F_{\mu\alpha}F_{\nu\beta}$, and $A\cdot A =
\bar{\gamma}^{\mu\nu} A_{\mu}A_{\nu}$. Hence $\bar{\gamma}^{\mu\nu}$ corresponds to
the canonical metric.
Now we wish to derive the  equations  from the effective action
(\ref{KKI2}). First of all, we have to change the non-linear term of
$\sr{|\delta^{\mu}_{\nu} + N^\mu_\nu|}$ with $N^\mu_\nu =\kappa^2 \Phi^3 A^{\mu}A_{\nu}$
into a manageable form.
 Considering  the small gauge coupling
 constant $(\kappa < 1)$, we assume that $\delta^{\mu}_{\nu}>
 N^\mu_\nu$. Using the formula as
 \beq
 \sr{\det[\bf 1 + \bf x]}= 1 + \fr{1}{2}{\bf tr}(\bf x) -\fr{1}{4}{\bf tr} (\bf x^2) +
 \fr{1}{8} ({\bf tr}(\bf x))^2+\cdots,
 \eeq
 one finds
 \beq
\sr{|\delta^{\mu}_{\nu} + N^\mu_\nu|}= 1+ \fr{1}{2}\kappa^2 \Phi^3 A \cdot A
-\fr{1}{8}(\kappa^2 \Phi^3 A \cdot A )^2+ \cdots.
\label{NONN}
\eeq
For simplicity we use here the relation of
\beq
\sr{|\delta^{\mu}_{\nu} + N^\mu_\nu|}
\simeq 1+ \fr{1}{2}\kappa^2 \Phi^3 A \cdot A.
\label{TNON}
\eeq
Even if we use
this instead of Eq.~(\ref{NONN}), we never lose the generality
for analyzing  the RS background.  To check it, we have the
relations
\beqa
\mbox{} && \fr{\delta \sr{|\delta^{\mu}_{\nu} + N^\mu_\nu|}}
{\delta \bar \gamma^{\mu\nu}} = \fr{1}{2}\kappa^2\Phi^3 A_{\mu}A_{\nu}
\big(1- \fr{1}{2} \kappa^2 \Phi^3 A \cdot A+ \cdots \big),   \label{DIFa}  \\
&& \fr{\delta \sr{|\delta^{\mu}_{\nu} + N^\mu_\nu|}}
{\delta \Phi} = \fr{3}{2} \kappa^2 \Phi^2A \cdot A
\big(1- \fr{1}{2} \kappa^2 \Phi^3 A \cdot A+ \cdots \big), \label{DIFb}  \\
&& \fr{\delta \sr{|\delta^{\mu}_{\nu} + N^\mu_\nu|}}
{\delta A^\mu} =\kappa^2\Phi^3A_{\mu}
\big(1- \fr{1}{2} \kappa^2 \Phi^3 A \cdot A+ \cdots \big).
\label{DIFc}
\eeqa

Making use of Eq.~(\ref{TNON}), we obtain the truncated equations of motion

\beqa
\mbox{} && R_{\mu\nu} = \fr{\kappa^2}{2}\Phi^3 \Big( F_{\mu\alpha}
  {F_{\nu}}^{\alpha} -\fr{1}{4}\bar{\gamma}_{\mu\nu}F^2 \Big)
  -6k^2\kappa^2 (1-\Phi )A_{\mu}A_{\nu}   \nonumber  \\
&& \qquad\qquad\qquad  +\fr{3}{2}\Phi^{-2}
  \Big[ \bar{\nabla}_{\mu}\Phi \bar{\nabla}_{\nu}\Phi
  -2k^2 \bar{\gamma}_{\mu\nu} (\Phi^{-1}+\Phi -2) \Big] ,
\label{EOMg}   \\
&& \bar{\nabla}^{\mu}F_{\mu\nu} +12k^2\kappa^2\Phi^{-3}(1-\Phi )A_{\nu}
  = -3\Phi^{-1} \bar{\nabla}^{\mu}\Phi F_{\mu\nu} ,
\label{EOMa}  \\
&& \bar{\nabla}^{\mu}\bar{\nabla}_{\mu} \Phi
  -\Phi^{-1}\bar{\nabla}^{\mu}\Phi \bar{\nabla}_{\mu}\Phi
  +2k^2 \Big[ \Phi^{-1}(4-\Phi -3\Phi^{-1}) -\kappa^2\Phi^2 A\cdot A
  \Big] = \fr{\kappa^2}{4} F^2 .
\label{EOMp}
\eeqa

It is easily checked  that $A_{\mu} =0$ and $\Phi =1$ satisfies
Eqs.~(\ref{EOMa}) and (\ref{EOMp}). Also if we use the full expression Eq.~(\ref{NONN}) for
 the non-linear term,
counting Eqs.~(\ref{DIFb}) and (\ref{DIFc}) with $ A_{\mu} =0$
leads to the same situation. In this case, Eq.~(\ref{EOMg})
leads to $R_{\mu\nu}=0$. Considering Eq.~(\ref{DIFa}) and (\ref{NONN}),
 we find $R_{\mu\nu}=0$.  Thus any 4D Ricci-flat metric
$\bar{\gamma}_{\mu\nu}$ is a solution to the 4D effective action Eq.~(\ref{KKI2}).
In particular,
$\bar{\gamma}_{\mu\nu} =\eta_{\mu\nu}= $diag$(-+++)$ corresponds to the RS
solution with Z$_2$-orbifold symmetry. It is noted that only for the case of $A_{\mu}
=0$, we can find the consistent solution. This is because the case
of $A_{\mu} \not=0$ results in the unwanted situation. That is, it
is not easy to find a solution to the equations including a lot of
terms like $ A \cdot A$, $(A \cdot A)^2,$ $\cdots$. This is why in
the RS approach they set $A_{\mu}=0$ at the beginning\cite{RS1}.

Now we are in a position to consider the perturbation analysis around the
RS ground state solution ($\bar{\gamma}_{\mu\nu}=\eta_{\mu\nu}, A_{\mu}=0,
\Phi =1$).  Let us introduce the  small fluctuations around the
RS solution
\beq
\gamma_{\mu\nu} = \eta_{\mu\nu} + \kappa h_{\mu\nu},
\qquad A_{\mu}=0 +a_{\mu}, \qquad  \Phi = 1+\kappa \varphi .
\eeq
Considering $g_{MN}= \eta_{MN} + \kappa h_{MN}$,  we have the
relations: $h_{5\mu}=-a_\mu$, $h_{55}= 2\varphi$.
We note that the non-zero $a_\mu$ breaks the Z$_2$-symmetry.
From Eq.(7), one finds
\beq
\bar{\gamma}_{\mu\nu} = \eta_{\mu\nu} +\kappa \bar{h}_{\mu\nu},
\qquad \bar{h}_{\mu\nu} = h_{\mu\nu} +\varphi \eta_{\mu\nu}.
\eeq
Then the bilinear action of Eq.~(\ref{KKI2})  which
governs the perturbative dynamics is given by~\cite{SSRSS,MC}
\beqa
I^{bil}_{zero-m} &=& \fr{\kappa^2}{16\pi G_4} \int d^4x \Big\{ -\fr{1}{4}\Big[
  \partial^{\mu}\bar{h}^{\alpha\beta}\partial_{\mu}\bar{h}_{\alpha\beta}
  -\partial^{\mu}\bar{h}\partial_{\mu}\bar{h} +2\partial^{\mu}\bar{h}_{\mu\nu}
  \partial^{\nu}\bar{h} -2\partial^{\mu}\bar{h}_{\mu\alpha}\partial^{\nu}
  {\bar{h}_{\nu}}^{\alpha} \Big]   \nonumber  \\
&& \qquad\qquad -\fr{1}{2}(\partial_{\mu}a_{\nu}\partial^{\mu}a^{\nu}
 -\partial_{\nu}a_{\mu}\partial^{\mu}a^{\nu})
  -\fr{3}{2}\partial_{\mu}\varphi \partial^{\mu}\varphi
  +6k^2 \varphi^2      \nonumber  \\
&& + \fr{1}{2} h_{\mu\nu}T^{\mu\nu} + a_\mu J^\mu +\varphi J_\varphi \Big\},
\label{KKILP}
\eeqa
where $\bar{h}=\eta^{\mu\nu}\bar{h}_{\mu\nu}=h+4\varphi$.
Here we introduce the 4D external sources of $(T^{\mu\nu}(x), J^\mu(x), J_\varphi(x))$
 to obtain the correct physical propagations.
Originally these all belong to the localized sources on the brane as
$(T^{\mu\nu}(x), J^\mu(x), J_\varphi(x))\delta(z)~\cite{GT,GKR}$.
 But after the integration over $z$ these lead to the last line of  Eq.~(\ref{KKILP}).
Surprisingly,
it turns out that although the Z$_2$-symmetry along $z$-axis is broken at the
linearized level,
the bilinear effective action
is invariant under the
U(1) gauge transformation.  Our previous observation about the U(1) symmetry
breaking  caused by the non-linear term and $A \cdot A$ is not correct at least
for the RS ground state solution. Here a nice combination of the non-linear term
and $A \cdot A$ in Eq.~(\ref{KKI2}) does not generate any mass term like $a \cdot a$.
This   appears as higher order term than the squared order:
 $6k^2\kappa^2 (1-\phi ) A\cdot A \  \to -6k^2\kappa^3 \varphi
a \cdot a$. We expect that this may
contribute to the quantum correction. However, if $\varphi=0$,
this term does not appear. The known method to obtain the
localization of the zero-modes is find the
bilinear action  without the external sources~\cite{BG,Oda1,Oda2}. The  action
Eq.~(\ref{KKILP}) obtained after the integration over $z$ and the
perturbation around the RS background is finite. Also it seems to have the canonical
forms for all fluctuation fields.  Hence, following
the conventional criterion, the zero-modes of the graviton,
graviscalar, graviphoton all are localized on the brane. However,
this  may be wrong  because it misses  the roles of the potential terms  and the external source.
The actual propagation of the physical zero-modes on the brane can be
justified by the calculation of their exchange amplitudes for the
sources\cite{DL,SSRSS,MC,KMPR}.

In order to understand what physical states there are, let us
derive
the linearized equations. From the action Eq.~(\ref{KKILP}) we
obtain
the equations of motion
\beqa
\mbox{} && \Box \bar{h}_{\mu\nu} +\partial_{\mu}\partial_{\nu}
  \bar{h} -\Big( \partial_{\mu}\partial^{\alpha}\bar{h}_{\alpha\nu}
  +\partial_{\nu}\partial^{\alpha}\bar{h}_{\alpha\mu} \Big)
  -\eta_{\mu\nu} \Big( \Box \bar{h} -\partial^{\alpha}\partial^{\beta}
  \bar{h}_{\alpha\beta} \Big) = -T_{\mu\nu} ,
\label{spin2}  \\
&& \Box a_{\mu} -\partial_{\mu} (\partial_{\nu}a^{\nu}) = -J_\mu,
\label{spin1}  \\
&& \Box \varphi +4k^2 \varphi = -\fr{1}{3} J_\varphi.
\label{spin0}
\eeqa
Here we find the 4D diffeomorphisms  plus the U(1) gauge symmetry.
Hence these can be taken into account by the source conservation
laws
\beq
\partial^\mu T_{\mu\nu}=0, \qquad\qquad \partial^\mu J_\mu=0.
\label{SCL}
\eeq
This means that the two equations Eqs.~(\ref{spin2}) and (\ref{spin1})
 are compatible with the source
conservation laws.
By taking the trace of Eq.~(\ref{spin2}), we have
\beq
\Box \bar{h} -\partial^{\alpha}\partial^{\beta} \bar{h}_{\alpha\beta}
  = \fr{1}{2} T
\label{Trace}
\eeq
with $T = T^\rho_\rho$.
Hence Eq.~(\ref{spin2}) becomes
\beq
\Box \bar{h}_{\mu\nu} +\partial_{\mu}\partial_{\nu}
  \bar{h} -\Big( \partial_{\mu}\partial^{\alpha}\bar{h}_{\alpha\nu}
  +\partial_{\nu}\partial^{\alpha}\bar{h}_{\alpha\mu} \Big)
  =-(T_{\mu\nu} - \fr{1}{2} \eta_{\mu\nu} T).
\label{spin2b}
\eeq

So far we have not chosen any gauge for $\bar{h}_{\mu\nu},a_\mu$. Now let us
choose the  harmonic  gauge  and Lorenz gauge, respectively
\beq
\partial^{\mu}\bar{h}_{\mu\nu} =\fr{1}{2}\partial_{\nu} \bar{h},
\qquad \partial_{\mu}a^{\mu} =0.
\eeq
 Using these gauge conditions, Eq.~(\ref{spin2b}) and
Eq.~(\ref{spin1}) reduce to
\beq
\Box \bar{h}_{\mu\nu} =-(T_{\mu\nu} - \fr{1}{2} \eta_{\mu\nu} T), \qquad
\Box a_{\mu} = - J_\mu.
\label{sp12}
\eeq
The first equation is exactly the same equation which was derived
for the graviton zero-mode using the brane-bending effect~\cite{GT}.
In the brane-bending calculation, it requires $h=0$.
Furthermore we get from Eq~.(\ref{Trace}) and (\ref{sp12}) the
trace equation
\beq
\Box \bar{h} = T.
\label{trah}
\eeq

To obtain the on-shell exchange amplitude induced by the sources,
let us plug Eqs.~(\ref{spin2}), (\ref{spin1}), and (\ref{spin0})
into Eq.~(\ref{KKILP})~\cite{SSRSS}. Then we find
\beq
I^{ampx}_{zero-m} = \fr{\kappa^2}{32\pi G_4} \int d^4x \Big\{
 \fr{1}{2} \bar h_{\mu\nu}(x)T^{\mu\nu}(x) + a_\mu(x) J^\mu(x) +\varphi(x)
 J_\varphi(x)
 \Big\}.
\label{AMPX}
\eeq
We wish to take its Fourier-transformed form which  makes the
calculation easy~\cite{DL},
\beq
I^{ampp}_{zero-m} = \fr{\kappa^2}{32\pi G_4} \int d^4p \Big\{
 \fr{1}{2} \bar h_{\mu\nu}(p)T^{\mu\nu}(p) + a_\mu(p) J^\mu(p) +\varphi(p)
 J_\varphi(p)
 \Big\}.
\label{AMPP}
\eeq
From Eqs.~(\ref{sp12}) and (\ref{spin0}), their Fourier-transformed fluctuations
 are given by
\beqa
\mbox{} && \bar{h}_{\mu\nu}(p)= \fr{1}{p^2}\big[ T_{\mu\nu}(p)-
\fr{1}{2}\eta_{\mu\nu} T(p) \big] ,
\label{spinp2}  \\
&&  a_{\mu}(p) = \fr{J_\mu(p)}{p^2},
\label{spinp1}  \\
&& \varphi(p) = \fr{J_\varphi(p)}{3(p^2 +m^2_\varphi)}
\label{spinp0}
\eeqa
with $m^2_\varphi= -4k^2$.
Substituting these into Eq.~(\ref{AMPP}) leads to
\beq
I^{ampp}_{zero-m} = \fr{\kappa^2}{32\pi G_4} \int d^4p \Big\{
 \fr{1}{2p^2} \Big(T^{\mu\nu}(p)T_{\mu\nu}(p)-\fr{1}{2}T^2(p)\Big) +
 \fr{J^\mu(p)J_\mu(p)}{p^2}  +
 \fr{J^2_\varphi(p)}{3(p^2 +m^2_\varphi)}
 \Big\}.
\label{AMPPS}
\eeq

In order to study the massless states, it is best to choose the
light-cone frame of $p_\mu=(p_1,0,0,p_4)$~\cite{SSRSS}. Then the source
conservation law of $p^\mu T_{\mu\nu}=0$ in Eq.~(\ref{SCL})
give us the relations
\beq
T_{11}=T_{41}=T_{44},~~ T_{12}=T_{42}, ~~T_{13}=T_{43}.
\label{SCL1}
\eeq
Using this, one finds the spin-2 exchange amplitude
\beq
T^{\mu\nu}(p)T_{\mu\nu}(p)-\fr{1}{2}T^2(p)=|T_{+2}|^2 +
|T_{-2}|^2,
\label{SCL2}
\eeq
where $T_{\pm2}=\fr{1}{2}(T_{22}-T_{33}) \pm i T_{23}$.
On the other hand, the source conservation law of  $p^\mu J^\mu=0$
leads to
\beq
J_1=J_4
\label{SCL3}.
\eeq
Making use of this, one has  the spin-1 exchange amplitude
\beq
2 J^\mu J_\mu = |J_{+1}|^2 + |J_{-1}|^2
\label{SCL4}
\eeq
with $J_{\pm1}=J_2 \pm i J_3$.
Finally, plugging this information into Eq.~(\ref{AMPPS}) leads to

\beq
I^{ampp}_{zero-m} = \fr{\kappa^2}{62\pi G_4} \int d^4p \Big\{
 \fr{1}{p^2} \Big(|T_{+2}|^2 +
|T_{-2}|^2 + |J_{+1}|^2 + |J_{-1}|^2 \Big) +
 \fr{2|J_\varphi|^2}{3(p^2 +m^2_\varphi)}
 \Big\},
\label{AMPPF}
\eeq
indicating that a total of four massless states: the gravitons
with helicities $\lambda=\pm 2$ and the graviphotons with
helicities $\lambda=\pm 1$~\cite{SSRSS}.
Therefore, it proves that $\bar{h}_{\mu\nu}$ and
$a_{\mu}$ indeed represent the massless spin-2 propagations and the
massless spin-1 propagations on the brane, respectively.
Of course these all are localized on the brane .

On the other hand, the graviscalar  $\varphi$ in
Eq.~(\ref{AMPPF}) appears to be massive. Unfortunately, it has the tachyonic
mass $m^2_{\varphi} = -4k^2$.
It may indicate
 the unstable, massless mode
on the brane. This means that the graviscalar cannot be
localized on the brane. This can be easily read off from the 4D
effective action in Eq.~(\ref{KKI1}).
Essentially we wish to treat the graviscalar  as a massless freedom.
However, the brane with tension $\sigma \sim k$ generates the
unwanted tachyonic terms for the graviscalar $\Phi$ which are proportional to
$k^2~$\footnote{ This  may imply
that our ansatz of $\gamma_{\mu\nu}(x), A_{\mu}(x), \Phi(x)$ is not appropriate for
integrating out the massive KK modes. We thank  Kakushadze for pointing out this.}.
In the absence of the sources, the consistency between the
linearized equation with $h=0$ leads to $\varphi=0$~\cite{ML}.
At this stage we comment on the role of the trace of $ h_{\mu\nu}$.
We assume that in the presence of the sources this is
the pure gauge degree of freedom, as was  in the brane-bending approach\cite{GT}.
Therefore we can choose
$ h =0$, which implies that  Eq.~(\ref{trah}) leads to $\Box \varphi=
T/4$. This leads to the contradiction to the graviscalar equation
of Eq.~(\ref{spin0}). We have two
equations (that is, one is massless, whereas the other is massive)
 for  the same field $\varphi$.
 Hence we  can choose $\varphi=0$ on the brane  for the consistency.

In conclusion we have established the localization of the
graviphotons on the brane. Although
these  belong to asymmetric modes which do not satisfy the
Z$_2$-background symmetry, they possess U(1)
symmetry. Also  the graviscalar
is not a massless scalar field  localized on the brane even for introducing
the matter sources.

\section*{Acknowledgments}

The author thanks Gungwon Kang and Hyungwon Lee  for helpful discussions.
This work was supported in part by the Brain Korea 21 Programme,  Project No. D-0025 and
KOSEF, Project No. 2000-1-11200-001-3.

\end{document}